\documentclass{article}
\usepackage{graphicx,cite,bm,amssymb}
\newcommand{\PRL}{\textit{Phys. Rev. Lett. }}
\newcommand{\PRA}{\textit{Phys. Rev. A }}

\newcommand{\PRE}{\textit{Phys. Rev. E }}

\newcommand{\JCS}{\textit{J. Cell Sci. }}
\newcommand{\JPA}{\textit{J. Phys. A: Math. Gen. }}
\newcommand{\bequ}{\begin{equation}}
\newcommand{\beqn}{\begin{eqnarray}}
\newcommand{\eequ}{\end{equation}}
\newcommand{\eeqn}{\end{eqnarray}}

\begin{document}
\title{Elasticities and stabilities: lipid membranes vs cell
membranes}
\author{Z. C. Tu$^1$, R. An$^2$, and Z. C. Ou-Yang$^1$\\
 $^1$\textit{Institute of Theoretical Physics,
 Chinese Academy of Sciences}\\
 \textit{P. O. Box 2735 Beijing 100080, China}\\
  $^2$\textit{Department of Mathematics}\\
\textit{Hong Kong University of Science and Technology}\\
\textit{Clear Water Bay, Kowloon, Hong Kong}} \date{}\maketitle

\section*{ABSTRACT}
A cell membrane can be simply regarded as composite material
consisting of lipid bilayer, membrane cytoskeleton beneath lipid
bilayer, and proteins embedded in lipid bilayer and linked with
membrane cytoskeleton if one only concerns its mechanical
properties. In this Chapter, above all, the authors give a brief
introduction to some important work on mechanical properties of
lipid bilayers following Helfrich's seminal work on spontaneous
curvature energy of lipid bilayers. Next, the entropy of a polymer
confined in a curved surface and the free energy of membrane
cytoskeleton are obtained by scaling analysis. It is found that
the free energy of cell membranes has the form of the in-plane
strain energy plus Helfrich's curvature energy. The equations to
describe equilibrium shapes and in-plane strains of cell membranes
by osmotic pressures are obtained by taking the first order
variation of the total free energy containing the elastic free
energy, the surface tension energy and the term induced by osmotic
pressure. The stability of spherical cell membrane is discussed
and the critical pressure is found to be much larger than that of
spherical lipid bilayer without membrane cytoskeleton. Lastly, the
authors try to extend the present static mechanical model of cell
membranes to the cell structure dynamics by proposing a group of
coupling equations involving tensegrity architecture of
cytoskeleton, fluid dynamics of cytoplasm and elasticities of cell
membranes.

\section*{INTRODUCTION}
Cells are the basic elements of living organisms, such as plants
and animals. Viewed from structure, an eukaryotic cell consists of
cell membrane, cytoplasm, cytoskeleton, nucleus and so on. Cell
membrane defines the boundary between the living cell and its
environment, including extracellular matrix and liquid
surroundings. Cytoplasm is a kind of viscous fluid. Cytoskeleton
is a three-dimensional structure composed of three classes of
fibers \cite{Lodish}: microtubules (20 nm in diameter), elements
built of polymers of the protein tubulin; microfilaments (7 nm in
diameter), built of the protein actin; and intermediate filaments
(10 nm in diameter), built of one or more rod-like protein
subunits. Nucleus of the eukaryotic cell is enclosed by membrane.

Membranes consist of lipids, proteins and carbohydrates etc.
Lipids and proteins are dominant components of membranes. One of
the principal types of lipids in membranes is phospholipid. A
phospholipid molecule has a polar hydrophilic head group and two
hydrophobic hydrocarbon tails. In physical point of view, a lipid
molecule can be simply regarded as an amphipathic rod. When a
quantity of lipid molecules disperse in water, they will assemble
themselves into a bilayer in which the hydrophilic heads shield
the hydrophobic tails from the water surroundings because of the
hydrophobic forces.

There are many simplified models for cell membranes in history
\cite{Edidin}. Among them, the widely accepted one is the fluid
mosaic model proposed by Singer and Nicolson in 1972
\cite{nicolson}. In their model, cell membrane is considered as a
lipid bilayer where the lipid molecules can move freely in the
membrane surface like fluid, while the proteins are embedded in
the lipid bilayer. Some proteins are called integral membrane
proteins because they traverse entirely in the lipid bilayer and
play the role of information and matter communications with both
the inside of the cell and its outer environment. The others are
called peripheral membrane proteins because they are partially
embedded in the bilayer and accomplish the other biological
functions. Beneath the lipid membrane, the membrane cytoskeleton,
a network of proteins, links with the proteins in lipid membrane.

The first step to study the elasticity of cell membrane is to
study lipid bilayer. Usually, the thickness of lipid bilayer is
about 4 nanometers which is much less than the scale of the whole
lipid bilayer (about several micrometers). Therefore, it is
reasonable to describe the lipid bilayer by a mathematical
surface. In 1973, Helfrich \cite{Helfrich} recognized that the
lipid bilayer was just like a nematic liquid crystal at room
temperature. Based on the elastic theory of liquid crystal
\cite{gennes}, he proposed the curvature energy per unit area of
the bilayer
\begin{equation}\label{Helfrich} f_c=(k_c/2)(2H+c_0)^2, \end{equation}
where $k_c$ is an elastic constant; and $H$, $c_0$ are mean
curvature and spontaneous curvature of the membrane surface,
respectively. In (\ref{Helfrich}), Gaussian curvature $K$ is not
written explicitly because only close bilayer is discussed in this
chapter. We can safely ignore the thermodynamic fluctuation of the
curved bilayer at the room temperature because of $k_c\approx
10^{-19}J\gg k_BT$ \cite{Duwe,Mutz2}, where $k_B$ is the Boltzmann
factor and $T$ the room temperature. Based on Helfrich's curvature
energy, we can express the free energy of a closed bilayer under
the osmotic pressure $\Delta p$ (the outer pressure minus the
inner one) as:
\begin{equation}\label{free-e-closed} \mathcal{F}_H=\int (f_c+\mu) dA+\Delta p\int
dV, \end{equation} where $dA$ is the area element and $V$ the
volume enclosed by the closed bilayer. $\mu$ is the surface
tension of the bilayer. Based on above free energy, many
researchers studied the shapes of bilayers
\cite{oy1,Reinhard,Seifertap}. Especially, by taking the first
order variation of above free energy and doing the complicated
calculations of tensors, Ou-Yang \emph{et al.} derived an equation
to describe the equilibrium shape of the bilayer as
\cite{oy2,oy2pra}:
\begin{equation}\label{shape-closed}\Delta p-2\mu
H+k_c(2H+c_0)(2H^2-c_0H-2K)+k_c\nabla^2(2H)=0. \end{equation} This
equation is now called the shape equation of closed membranes or
generalized Laplace equation. They also obtained that the
threshold pressure for instability of spherical bilayer was
$\Delta p_{c}\sim k_c/R^3$, where $R$ being the radius of
spherical bilayer.

Using the shape equation (\ref{shape-closed}) of closed bilayers,
Ou-Yang predicted that there was a lipid torus with the ratio of
two radii being exactly $\sqrt{2}$ \cite{oy4}. His prediction was
soon confirmed by the experiments \cite{Mutz,Linz,Rudolph}.
Otherwise, researchers obtained an analytical solution to
Eq.(\ref{shape-closed}) which explained the classical
problem\cite{fengyz}---the biconcave discoidal shape of normal red
cells \cite{oy3}.

Recently, Tu and Ou-Yang have proposed a mathematical scheme to
discuss the elasticities and stabilities of cell membranes with
membrane cytoskeleton and found that the membrane cytoskeleton
enhances the stabilities of cell membranes \cite{tzcjpa}. But they
have omitted the effect of in-plane modes of membranes on the
stabilities. In this chapter, we will first retrospect to
elasticities and stabilities of lipid bilayers following
Helfrich's seminal work on spontaneous curvature energy of lipid
bilayers. Next, we will fully discuss the entropic elasticity of
membrane cytoskeleton, as well as the elasticities and stabilities
of cell membranes with membrane cytoskeleton. Last, we will
expatiate on how to construct the framework of cell structure
dynamics involving tensegrity architecture of cell cytoskeleton,
fluid dynamics of cytoplasm and elasticities of cell membranes
(with membrane cytoskeleton).

\section*{LIPID MEMBRANE}
In this section, we will recur to some main results on
elasticities and stabilities of lipid bilayers by adopting the
mathematical scheme proposed in Ref.\cite{tzcjpa,tzcpre,antu}.

\begin{figure}[!htp]
\begin{center}
\includegraphics[width=8cm]{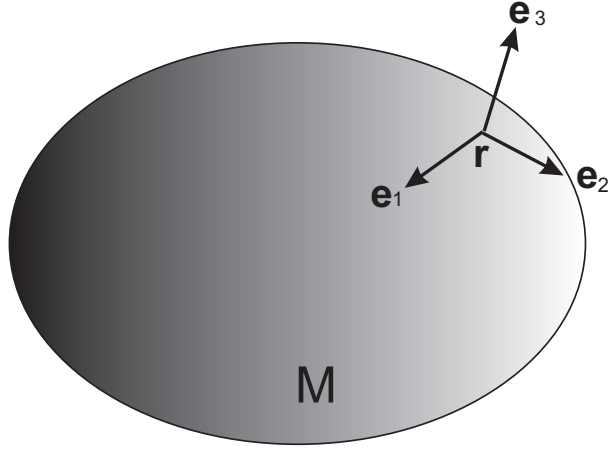}
\caption{\label{surfacem} Smooth and orientable surface $M$. we
can construct a right-hand orthonormal system
$\{\mathbf{e}_1,\mathbf{e}_2,\mathbf{e}_3\}$ at any point
$\mathbf{r}$ in the surface and call
$\{\mathbf{r};\mathbf{e}_1,\mathbf{e}_2,\mathbf{e}_3\}$ a moving
frame.}\end{center}
\end{figure}

We use a smooth and closed surface $M$ in 3-dimensional Euclid
space $\mathbb{E}^3$ to represent a membrane. As shown in
Fig.~\ref{surfacem}, we can construct a right-hand orthonormal
system $\{\mathbf{e}_1,\mathbf{e}_2,\mathbf{e}_3\}$ at any point
$\mathbf{r}$ in the surface and call
$\{\mathbf{r};\mathbf{e}_1,\mathbf{e}_2,\mathbf{e}_3\}$ a moving
frame. The differential of the frame is denoted by
\begin{equation}\label{infiniter}\left\{
\begin{array}{l}
d\mathbf{r}=\omega_1\mathbf{e}_1+\omega_2\mathbf{e}_2,\\
\label{dei}d\mathbf{e}_i=\omega_{ij}\mathbf{e}_j\quad (i=1,2,3),
\end{array}\right.\end{equation} where $\omega_1$, $\omega_2$ and $\omega_{ij}=-\omega_{ji}$
$(i,j=1,2,3)$ are 1-forms. The structure equations of the surface
are \beqn \label{domega1}
d\omega_1&=&\omega_{12}\wedge\omega_2;\\
\label{domega2}
d\omega_2&=&\omega_{21}\wedge\omega_1;\\
\label{omega13} \omega_{13}&=&a\omega_{1}+b\omega_{2},\quad \omega_{23}=b\omega_{1}+c\omega_{2};\\
\label{domgaij} d\omega_{ij}&=&\omega_{ik}\wedge\omega_{kj}\quad
(i,j=1,2,3). \eeqn Readers should notice that the operator ``$d$"
is an exterior differential operator \cite{tzcjpa} in this
chapter. The area element, mean curvature and Gaussian curvature
are respectively expressed as \cite{chen2}: \beqn
&&dA=\omega_1\wedge\omega_2, \\
&&H=(a+c)/2 \label{meanh},\\
&&K=ac-b^2\label{gassianK}. \eeqn

If $M$ undergoes an infinitesimal deformation such that every
point $\mathbf{r}$ of $M$ has a displacement $\delta\mathbf{r}$,
we obtain a new surface
$M'=\{\mathbf{r}'|\mathbf{r}'=\mathbf{r}+\delta\mathbf{r}\}$.
$\delta\mathbf{r}$ is called the variation of surface $M$ and can
be expressed as \beqn \delta
\mathbf{r}=\delta_1\mathbf{r}+\delta_2\mathbf{r}+\delta_3\mathbf{r},\label{deltar}\\
\delta_i\mathbf{r}=\Omega_{i}\mathbf{e}_{i}\quad
(i=1,2,3),\label{deltari} \eeqn where the repeated subindexes do
not represent Einstein summation. Due to the deformation of $M$,
${\mathbf{e}_1,\mathbf{e}_2,\mathbf{e}_3}$ also change. We denote
the change as \bequ \delta_l \mathbf{e}_{i}=\Omega
_{lij}\mathbf{e}_{j},\quad \Omega _{lij}=-\Omega _{lji}.\eequ

Using the commutativity between $\delta_i$ $(i=1,2,3)$ and $d$, we
obtain the fundamentally variational identities of the move frame
\cite{tzcjpa}: \beqn
&&\delta_1 \omega _{1} =d\Omega _{1}-\omega _{2}\Omega _{121},\label{omvaratione11}\\
&&\delta_1 \omega _{2} =\Omega _{1}\omega _{12}-\omega _{1}\Omega _{112},\label{omvaratione12}\\
&&\Omega _{113}=a\Omega _{1},\quad\Omega _{123}=b\Omega
_{1};\label{omvaratione13}\\
&&\delta _{2}\omega _{1} =\Omega _{2}\omega _{21}-\omega _{2}\Omega _{221},\label{omvaratione21} \\
&&\delta _{2}\omega _{2} =d\Omega _{2}-\omega _{1}\Omega _{212},\label{omvaratione22} \\
&&\Omega _{213} =b\Omega _{2},\quad\Omega _{223}=c\Omega
_{2};\label{omvaratione23}\\
&&\delta_3 \omega _{1} =\Omega _{3}\omega _{31}-\omega _{2}\Omega
_{321},
\label{detaomega1} \\
&&\delta_3 \omega _{2} =\Omega _{3}\omega _{32}-\omega _{1}\Omega
_{312},
\label{detaomega2} \\
&&d\Omega _{3} =\Omega _{313}\omega _{1}+\Omega _{323}\omega _{2};
\label{domega3}\\
&&\delta_l \omega _{ij}=d\Omega _{lij}+\Omega _{lik}\omega
_{kj}-\omega _{ik}\Omega _{lkj}.\label{detaomegaij}\eeqn

Now we can take the first order variation of functional
(\ref{free-e-closed}) by considering above variational identities
(\ref{omvaratione11})--(\ref{detaomegaij}) and Stokes theorem. It
is not hard to obtain \cite{tzcjpa}: \beqn \delta\mathcal{F}_H
=\int[\Delta p-2\mu
H&+&k_c(2H+c_0)(2H^2-c_0H-2K)\nonumber\\
&+&k_c\nabla^2(2H)]\Omega_3 dA.\label{deltafunc}\eeqn Thus the
Euler-Lagrange equation corresponding to the functional
$\mathcal{F}_H$ reduces to equation (\ref{shape-closed}) because
$\Omega_3$ is an arbitrary function. Equation (\ref{shape-closed})
describes the equilibrium shapes of lipid membranes.

Similarly, we can obtain the second order variation of functional
(\ref{free-e-closed}): \beqn \delta^2 \mathcal{F}_H&=&\int\Omega
_{3}\nabla ^{2}\Omega _{3}\left[
k_{c}(14H^{2}+2Hc_{0}-4K-c_{0}^{2}/2)-\mu \right]
dA+k_{c}\int(\nabla
^{2}\Omega _{3})^{2}dA \nonumber\\
&+&\int k_{c}(2H+c_{0})\left[ \nabla (2H\Omega _{3})\cdot \nabla
\Omega _{3}-2\nabla \Omega _{3}\cdot \tilde{\nabla}\Omega
_{3}-4\Omega _{3}\nabla
\cdot \tilde{\nabla}\Omega _{3}\right] dA\nonumber \\
&+&\int \Omega _{3}^{2}\left[
k_{c}(16H^{4}-20H^{2}K+4K^{2}+Kc_{0}^{2})+2K\mu -2H\Delta p\right]
dA.\label{secondvarnoK} \eeqn The detailed expressions of
operators $\nabla$, $\bar{\nabla}$, $\tilde{\nabla}$, $\nabla^2$,
$\nabla\cdot\bar{\nabla}$, $\nabla\cdot\tilde{\nabla}$ in above
equation can be found in Appendix D of Ref.\cite{tzcjpa}.

Now, we discuss mechanical stability of spherical lipid membrane
with radius $R$. In this case, we have $H=-1/R$ and $K=1/R^2$.
Substituting them into equation (\ref{shape-closed}), we arrive at
\bequ \Delta pR^2+2\mu
R-k_cc_0(2-c_0R)=0.\label{sphericalbilayer}\eequ Using
$\mathbf{r}=R(\sin\theta\cos\phi,\sin\theta\sin\phi,\cos\theta)$
to represent the point in the spherical membrane, we have
$\tilde{\nabla}=-(1/R)\nabla$, $\nabla \cdot
\tilde{\nabla}=-(1/R)\nabla^2$ and
$\nabla^2=\frac{1}{R^2\sin\theta}\frac{\partial}{\partial\theta}
\left(\sin\theta\frac{\partial}{\partial\theta}\right)
+\frac{1}{R^2\sin^2\theta}\frac{\partial^2}{\partial\phi^2}$.
Under the condition (\ref{sphericalbilayer}), equation
(\ref{secondvarnoK}) is transformed into \beqn \delta
^{2}\mathcal{F}_H&=&(2c_{0}k_{c}/R+\Delta pR)\int_{0}^{\pi }\sin
\theta d\theta \int_{0}^{2\pi }d\phi \Omega
_{3}^{2}\nonumber\\&+&(k_{c}c_{0}R+2k_{c}+\Delta
pR^{3}/2)\int_{0}^{\pi }\sin \theta d\theta \int_{0}^{2\pi
}d\phi\Omega _{3}\nabla ^{2}\Omega
_{3}\nonumber\\&+&k_{c}R^{2}\int_{0}^{\pi }\sin \theta d\theta
\int_{0}^{2\pi }d\phi (\nabla ^{2}\Omega
_{3})^{2}.\label{secondvsph}\eeqn Expand $\Omega _{3}$ by the
spherical harmonic functions \cite{wangzx}: \bequ\Omega _{3}
=\sum_{l=0}^{\infty
}\sum_{m=-l}^{m=l}a_{lm}Y_{lm}(\theta,\phi),\quad a_{lm}^{*}
=(-1)^{m}a_{l,-m}.\label{harmonicf}\eequ If considering $\nabla
^{2}Y_{lm} =-l(l+1)Y_{lm}/R^{2}$ and $\int_{0}^{\pi}\sin \theta
d\theta\int_{0}^{2\pi}d\phi Y_{lm}^{*}Y_{l^{\prime}m^{\prime}}
=\delta _{mm^{\prime }}\delta _{ll^{\prime}}$, we transform
equation (\ref{secondvsph}) into \bequ\delta^{2}\mathcal{F}_H=
(R/2)\sum_{l,m}|a_{lm}|^{2}[l(l+1)-2]\{2k_{c}/R^{3}[l(l+1)-c_{0}R]-\Delta
p\}.\eequ

Obviously, $\delta^{2}\mathcal{F}_H$ is a positive definite form
if $\Delta p<p_l\equiv (2k_{c}/R^{3})[l(l+1)-c_{0}R]\quad
(l=2,3,\cdots)$. Therefore, we can take the critical pressure as
\bequ \Delta
p_c=\min\{p_l\}=p_2=(2k_{c}/R^{3})(6-c_{0}R).\label{criticalpsc}\eequ
If $\Delta p>\Delta p_c$, the spherical bilayer will be instable
and inclined to transform into the biconcave discoid shape.

In this section, we have discussed the elasticity of and stability
of lipid bilayer under the pressure. But lipid bilayer is
oversimplified model of cell membrane which contains the membrane
cytoskeleton. That is, for cell membrane, we must take into
account the contribution of membrane cytoskeleton.

\section*{MEMBRANE CYTOSKELETON}
In this section, we will discuss the contribution of membrane
cytoskeleton to the free energy of cell membrane. The membrane
cytoskeleton is cross-linking chain-like protein structure which
can be thought of as a polymer membrane. Now we will derive its
free energy by analogy with the polymer membrane \cite{tuge}.

We take de Gennes' convention \cite{gennes2} in the this section:
the entropy $S$ is a dimensionless quantity and Boltzmann factor
$k_B$ is implicated in temperature $T$. If we regard the protein
in membrane cytoskeleton as Gaussian chains \cite{Dio}, its root
of mean square end-to-end distance is $R_0\sim \sqrt{N}b_0$, where
$b_0$ is the segment length of protein and $N\gg 1$ is the number
of segments. Assume that the principal radii of the membrane are
much larger than $R_0$. If we denote the in-plane strain tensor by
${\bm\epsilon}$ which is assumed to be a small quantity, the
entropy of the protein chain must be the function of $2HR_0$,
$KR_0^2$, $2J$ and $Q$ because entropy is a dimensionless
invariable under the transformation of coordinates, where $H$,
$K$, $J=\mathrm{tr}{\bm\epsilon}$ and $Q=\mathrm{tr}{\bm\epsilon}$
are the mean curvature, the Gaussian curvature, the trace of
strain tensor and the determinant of strain tensor, respectively.
Thus we can expand it as $S\sim
A_1(2HR_0)+A_2(2HR_0)^2+A_3KR_0^2+B_2(2J)^2+B_3Q$ up to the second
order terms, where $A_1, A_2, A_3, B_2, B_3$ are constants. There
is no first order term of $2J$ in the expression of the entropy
because we expect that $-{\bm\epsilon}$ plays the same role as
${\bm\epsilon}$ in the entropy. It is useful to write the entropy
in another equivalent form $S \sim
A_{2}R_{0}^{2}(2H+c_{0})^{2}+A_{3}R_{0}^{2}K+B_{2}(2J)^{2}+B_{3}Q$,
 where $c_0$ is a constant called spontaneous curvature and we
expect $|c_0R_0|\ll 1$. Consequently, the elastic free energy per
area of a membrane consisting of protein chains has the following
form
\begin{equation}\label{elastic}
f=-(Mh)TS=\frac{k_{d}}{2}[(2J)^{2}-\nu
Q]+\frac{k'_{c}}{2}(2H+c_{0})^{2}-\bar{k}'K,
\end{equation}
where $h$ is thickness of the membrane, $M$ the number of protein
chains per volume, and $k_d$, $\nu$, $k'_c$, $\bar{k}'$ are
unknown universal constants. Here we neglect the entanglement of
proteins. We will show $k_d=4MhT$ and $\nu=1$ as follows.

Let us consider an ideal case--the planar membrane with the
homogenous in-plane strains. In this case, $H$, $K$ and $c_0$ are
vanishing for planar membrane with symmetry between its two sides.
On the one hand, equation (\ref{elastic}) is simplified as
\begin{equation}\label{planar}
f=\frac{k_{d}}{2}[(2J)^{2}-\nu Q].
\end{equation}
For homogenous stain ${\bm \epsilon}$, we can express it by its
components $\epsilon_{11}$, $\epsilon_{22}$ and
$\epsilon_{12}=\epsilon_{21}=0$ in some orthonormal coordinate
system so that $2J=\epsilon_{11}+\epsilon_{22}$ and
$Q=\epsilon_{11}\epsilon_{22}$.

On the other hand, this structure can be compared with the
structure of rubber. In terms of the elasticity theory of rubber
\cite{Treloar}, the deformation energy of a planar rubber per area
can be expressed as
$f_r=(MhT/2)[\lambda_1^2+\lambda_2^2+1/(\lambda_1^2\lambda_2^2)-3]$,
where $\lambda_1=1+\epsilon_{11}$ and $\lambda_2=1+\epsilon_{22}$
are extensions. For small strains, it is expanded to the lowest
order terms as $f_r\sim
2MhT(\epsilon_{11}^2+\epsilon_{11}\epsilon_{22}+\epsilon_{22}^2)=2MhT[(2J)^2-Q]$.
Thus we can obtain $k_d=4MhT$ and $\nu=1$ by comparing it with
equation (\ref{planar}).

In this section we do not discuss the mechanical property of
membrane cytoskeleton alone. We will discuss it with the cell
membrane in the next section.

\section*{CELL MEMBRANE WITH MEMBRANE CYTOSKELETON}
The free energy of cell membrane with membrane cytoskeleton is
taken as the sum of the free energies of lipid bilayer and
membrane cytoskeleton. Thus we write the free energy of cell
membrane under the osmotic pressure $\Delta p$ as:
\bequ\label{freeeng}\mathcal{F}=\int (k_{d}/2)[(2J)^{2}-Q]dA+\int
[(\bar{k}_{c}/2) (2H+\bar{c}_{0})^{2}+\mu ]dA+\Delta p\int
dV,\eequ where $\bar{k}_{c}=k_c+k'_c$,
$\bar{c}_{0}=k_cc_0/\bar{k}_c$, and the term related to the
Gaussian curvature disappears because its integration $\int KdA$
is an unimportant constant so that it is omitted.

Before taking the first order variation of the free energy
(\ref{freeeng}), we must introduce the strain analysis expressed
by the notation of differential forms \cite{tzcjpa}.

If a point $\textbf{r}_0$ in a surface undergoing a displacement
$\textbf{u}$ to arrive at point $\textbf{r}$, we have $d\mathbf{u}
=d\mathbf{r}-d\mathbf{r}_{0}$ and naturally $\delta
_{i}d\mathbf{u} =\delta _{i}d\mathbf{r}$ ($i=1,2,3$).

If denote $d\mathbf{r}=\omega _{1}\mathbf{e}_{1}+\omega
_{2}\mathbf{e}_{2}$ and $d\mathbf{u} =\mathbf{U}_{1}\omega
_{1}+\mathbf{U}_{2}\omega _{2}$ with $|\mathbf{U}_{1}| \ll
1,|\mathbf{U}_{2}|\ll 1$, we can define the in-plane strains
\cite{wujk}: \beqn
\varepsilon _{11} &=&\left[ \frac{d\mathbf{u}\cdot \mathbf{e}_{1}}{|d\mathbf{%
r}_{0}|}\right] _{\omega _{2}=0}\approx \mathbf{U}_{1}\cdot \mathbf{e}_{1},\label{epsl11} \\
\varepsilon _{22} &=&\left[ \frac{d\mathbf{u}\cdot \mathbf{e}_{2}}{|d\mathbf{%
r}_{0}|}\right] _{\omega _{1}=0}\approx \mathbf{U}_{2}\cdot \mathbf{e}_{2},\label{epsl22} \\
\varepsilon _{12} &=&\frac{1}{2}\left[ \left( \frac{d\mathbf{u}\cdot \mathbf{%
e}_{2}}{|d\mathbf{r}_{0}|}\right) _{\omega _{2}=0}+\left( \frac{d\mathbf{u}%
\cdot \mathbf{e}_{1}}{|d\mathbf{r}_{0}|}\right) _{\omega
_{1}=0}\right]
\approx \frac{1}{2}\left( \mathbf{U}_{1}\cdot \mathbf{e}_{2}+\mathbf{U}%
_{2}\cdot \mathbf{e}_{1}\right).\label{epsl12} \eeqn

Using $\delta _{i}d\mathbf{u} =\delta _{i}d\mathbf{r}$ and the
definitions of strains (\ref{epsl11})--(\ref{epsl12}), we can
obtain the leading terms of variational relations: \beqn \delta
_{i}\varepsilon _{11}\omega _{1}\wedge \omega _{2} &=&\delta
_{i}\omega _{1}\wedge \omega _{2},\label{epsli11} \\
\delta _{i}\varepsilon _{12}\omega _{1}\wedge \omega _{2} &=&\frac{1}{%
2}[\omega _{1}\wedge \delta _{i}\omega _{1}+\delta _{i}\omega
_{2}\wedge
\omega _{2}],\label{epsli12} \\
\delta _{i}\varepsilon _{22}\omega _{1}\wedge \omega _{2}
&=&\omega _{1}\wedge \delta _{i}\omega _{2}.\label{epsli22} \eeqn

From equations (\ref{omvaratione11})--(\ref{detaomegaij}) and
(\ref{epsli11})--(\ref{epsli22}), we have \cite{tzcjpa}: \beqn
\delta _{1}\mathcal{F}&=&\int k_d[-d(2J)\wedge \omega
_{2}-\frac{\varepsilon _{11}d\omega _{2}-\varepsilon _{12}d\omega
_{1}}{2}+\frac{d(\varepsilon _{12}\omega _{1}+\varepsilon
_{22}\omega _{2})}{2}]\Omega
_{1},\label{deltamf1} \\
\delta _{2}\mathcal{F}&=& \int k_d[d(2J)\wedge \omega
_{1}-\frac{\varepsilon _{12}d\omega _{2}-\varepsilon _{22}d\omega
_{1}}{2}-\frac{d(\varepsilon _{11}\omega _{1}+\varepsilon
_{12}\omega _{2})}{2}]\Omega
_{2},\label{deltamf2}\\
\delta
_{3}\mathcal{F}&=&\int[\bar{k}_{c}(2H+\bar{c}_{0})(2H^{2}-\bar{c}_{0}H-2K)+\bar{k}_{c}\nabla
^{2}(2H)\nonumber\\&&\qquad+\Delta p-2H(\mu
+k_{d}J)-\frac{k_{d}}{2}(a\varepsilon _{11}+2b\varepsilon
_{12}+c\varepsilon _{22})]\Omega _{3}dA.\label{deltamf3} \eeqn
Thus the Euler-Lagrange equations corresponding to the functional
(\ref{freeeng}) are \beqn &&k_d[-d(2J)\wedge \omega
_{2}-\frac{1}{2}(\varepsilon _{11}d\omega _{2}-\varepsilon
_{12}d\omega _{1})+\frac{1}{2}d(\varepsilon _{12}\omega
_{1}+\varepsilon _{22}\omega _{2})] =0,\label{shapecm1} \\
&&k_d[d(2J)\wedge \omega _{1}-\frac{1}{2}(\varepsilon _{12}d\omega
_{2}-\varepsilon _{22}d\omega _{1})-\frac{1}{2}d(\varepsilon
_{11}\omega
_{1}+\varepsilon _{12}\omega _{2})] =0, \label{shapecm2}\\
&&\Delta p-2H(\mu
+k_{d}J)+\bar{k}_{c}(2H+\bar{c}_{0})(2H^{2}-\bar{c}_{0}H-2K)+\bar{k}_{c}\nabla
^{2}(2H)\nonumber\\&&\qquad-\frac{k_{d}}{2}(a\varepsilon
_{11}+2b\varepsilon _{12}+c\varepsilon _{22}) =0.
\label{shapecm3}\eeqn Equations (\ref{shapecm1}) and
(\ref{shapecm2}) are called the in-plane strain equations because
they describe the in-plane strains of cell membrane under the
pressure $\Delta p$. Equation (\ref{shapecm3}) is called the shape
equation because it describes the equilibrium shape of cell
membrane under the pressure $\Delta p$.

If we take a local orthonormal coordinates $(u^1,u^2)$ such that
the first and second fundamental forms of the surface are
$I=g_{11}(du^1)^2+g_{22}(du^2)^2$ and
$II=L_{11}(du^1)^2+2L_{12}du^1du^2+L_{22}(du^2)^2$, respectively,
above equations (\ref{shapecm1})--(\ref{shapecm3}) can be
expressed as: \beqn
k_d\left[(\epsilon _{22}-\epsilon _{11})\frac{\partial \sqrt{g_{22}}}{%
\partial u^{1}} -\sqrt{g_{22}}\frac{\partial }{\partial u^{1}}(2\epsilon
_{11}+\epsilon _{22})-\sqrt{g_{11}}\frac{\partial \epsilon _{12}}{%
\partial u^{2}}-2\epsilon _{12}\frac{\partial \sqrt{g_{11}}}{\partial
u^{2}}\right]&=&0, \nonumber\\
k_d\left[(\epsilon _{11}-\epsilon _{22})\frac{\partial \sqrt{g_{11}}}{%
\partial u^{2}}-\sqrt{g_{11}}\frac{\partial }{\partial u^{2}}(\epsilon
_{11}+2\epsilon _{22})-\sqrt{g_{22}}\frac{\partial \epsilon _{12}}{%
\partial u^{1}}-2\epsilon _{12}\frac{\partial \sqrt{g_{22}}}{\partial
u^{1}}\right] &=&0, \nonumber\\
\Delta p-2(\mu
+k_{d}J)H+\bar{k}_{c}(2H+\bar{c}_{0})(2H^{2}-\bar{c}_{0}H-2K)+\bar{k}_{c}\nabla
^{2}(2H) && \nonumber\\
-\frac{k_{d}}{2}[a\epsilon _{11}+2b\epsilon _{12}+c\epsilon _{22}]
&=&0.\nonumber \eeqn

Obviously, if $k_d=0$, then equations (\ref{shapecm1}) and
(\ref{shapecm2}) are two identities. Moreover equation
(\ref{shapecm3}) degenerates into shape equation
(\ref{shape-closed}) of closed lipid bilayers in this case.
Generally speaking, it is difficult to find the analytical
solutions to equations (\ref{shapecm1})--(\ref{shapecm3}). But we
can verify that $\epsilon_{11}=\epsilon_{22}=\varepsilon$ (a
constant), $\epsilon_{12}=0$ can satisfy
(\ref{shapecm1})--(\ref{shapecm3}) for a spherical membrane with
radius $R$ if the following equation is valid:
\begin{equation}\label{spherecem1}
\Delta pR^{2}+(2\mu
+3k_{d}\varepsilon)R+\bar{k}_{c}\bar{c}_{0}(\bar{c}_{0}R-2)=0.
\end{equation}

Now we discuss the effect of membrane cytoskeleton on the
stability of cell membrane. That is, we will calculate the second
order variation of functional (\ref{freeeng}). Additionally, we
only consider the spherical membrane with uniform in-plane strains
whose radius satisfies equation (\ref{spherecem1}) for simplicity.

In Ref.\cite{tzcjpa}, only the term $\delta _{3}^2\mathcal{F}$
related to the out-plane mode is calculated. Here we also consider
the contribution of in-plane modes. Due to the notation of
exterior differential and Hodge star $\ast$, $\Omega_1$ and
$\Omega_2$ can be expressed as $\Omega _{1}\omega _{1}+\Omega
_{2}\omega _{2}=d\Omega +\ast d\chi$ by two scalar functions
$\Omega$ and $\chi$. Using equations
(\ref{omvaratione11})--(\ref{detaomegaij}) and
(\ref{epsli11})--(\ref{epsli22}), we can calculate $\delta
_{1}^2\mathcal{F}$, $\delta _{2}^2\mathcal{F}$, $\delta
_{3}^2\mathcal{F}$, $\delta _{1}\delta _{2}\mathcal{F}$, $\delta
_{1}\delta _{3}\mathcal{F}$, and $\delta _{2}\delta
_{3}\mathcal{F}$ from equations (\ref{deltamf1})--(\ref{deltamf3})
and (\ref{spherecem1}). Eventually, we arrive at \bequ\delta
^2\mathcal{F}=\delta _{1}^2\mathcal{F}+\delta
_{2}^2\mathcal{F}+\delta _{3}^2\mathcal{F}+2\delta _{1}\delta
_{2}\mathcal{F}+2\delta _{1}\delta _{3}\mathcal{F}+2\delta
_{2}\delta _{3}\mathcal{F}\equiv G_1+G_2,\eequ where \beqn
G_1&=&\int\Omega
_{3}^{2}\{3k_{d}/R^{2}+(2\bar{k}_{c}\bar{c}_{0}/R^{3})+\Delta
p/R\}dA\nonumber
\\&&+\int\Omega
_{3}\nabla ^{2}\Omega
_{3}\{\bar{k}_{c}\bar{c}_{0}/R+2\bar{k}_{c}/R^{2}+\Delta
pR/2\}dA+\int \bar{k}_{c}(\nabla
^{2}\Omega _{3})^{2}dA\nonumber \\
&&+\frac{3k_{d}}{R}\int\Omega _{3}\nabla ^{2}\Omega
dA+k_{d}\int\left( \nabla ^{2}\Omega \right) ^{2}dA+\frac{k_{d}}{2R^{2}}%
\int\Omega \nabla ^{2}\Omega dA,\\
G_2&=&\frac{k_{d}}{4}\int(\nabla ^{2}\chi )^{2}dA+\frac{k_{d}}{2R^{2}}%
\int\chi \nabla ^{2}\chi dA. \eeqn If we take
$\kappa=\bar{k}_c/2$, $K=3k_d/2$, $\mu=k_d/2$, $w=\Omega_3$ and
$\Psi=\Omega$ in equations (6) and (7) of Zhang \textit{et al.}'s
paper \cite{zhangz}, then $G_1$ and $G_2$ correspond to
$F_1[w,\Psi]$ and $F_2[\chi]$ in that paper under the conditions
of $\Delta p=0$ and $\bar{c}_0R=2$. Obviously, there is no
coupling between modes $\{\chi\}$ and $\{\Omega,\Omega_3\}$; but
there is coupling between in-plane mode $\{\Omega\}$ and
out-of-plane mode $\{\Omega_3\}$. We will show that in-plane modes
have quantitive effect on the stability of the cell membrane
although they can not qualitatively modify the results of
Ref.\cite{tzcjpa}.

Because $G_2$ is obviously positive definite, we merely need to
discuss $G_1$. $\Omega_{3}$ and $\Omega$ in the expression of
$G_1$ can be expanded by spherical harmonic functions
\cite{wangzx} as $\Omega _{3} =\sum_{l=0}^{\infty
}\sum_{m=-l}^{m=l}a_{lm}Y_{lm}(\theta,\phi)$ and $\Omega
=\sum_{l=0}^{\infty }\sum_{m=-l}^{m=l}b_{lm}Y_{lm}(\theta,\phi)$
with $a_{lm}^{*} =(-1)^{m}a_{l,-m}$ and $b_{lm}^{*}
=(-1)^{m}b_{l,-m}$. It follows that \beqn G_{1}
&=&\sum_{l=0}^{\infty
}\sum_{m=0}^{l}2|a_{lm}|^{2}%
\{3k_{d}+[l(l+1)-2][l(l+1)\bar{k}_{c}/R^{2}-\bar{k}_{c}\bar{c}_{0}/R-\Delta pR/2]\} \nonumber\\
&&-\sum_{l=0}^{\infty
}\sum_{m=0}^{l}\frac{3k_{d}}{R}l(l+1)(a_{lm}^{\ast
}b_{lm}+a_{lm}b_{lm}^{\ast })\nonumber\\ &&+\sum_{l=0}^{\infty }\sum_{m=0}^{l}\frac{k_{d}}{%
R^{2}}\left[ 2l^{2}(l+1)^{2}-l(l+1)\right] |b_{lm}|^{2}. \eeqn

We find that if $\Delta p<p_{l}=\frac{3k_{d}}{\left[ 2l(l+1)-1\right] R}+\frac{2\bar{k}_{c}[l(l+1)-\bar{c}_{0}R]%
}{R^{3}}\quad(l=2,3,\cdots)$, then $G_{1}$ is positive definite,
i.e., the membrane is stable. We must take the minimum of $p_l$ to
obtain the critical pressure:
\begin{equation}\Delta p_{c}=\min \{p_{l}\}=\left\{
\begin{array}{c}
\frac{3k_{d}}{11R}+\frac{2\bar{k}_{c}[6-\bar{c}_{0}R]}{R^{3}}<\frac{\bar{k}_{c}[23-2\bar{c}_{0}R]}{%
R^{3}},\quad (3k_{d}R^{2}<121\bar{k}_{c}) \\
\frac{2\sqrt{3k_{d}\bar{k}_{c}}}{R^{2}}+\frac{\bar{k}_{c}}{R^{3}}(1-2\bar{c}_{0}R),\quad
(3k_{d}R^{2}>121\bar{k}_{c})%
\end{array}%
\right.. \label{criticalpcm}
\end{equation}

But if we do not consider the in-plane mode $\{\Omega\}$, we will
obtain the critical pressure \cite{tzcjpa}:\bequ\Delta
p_{c}=\left\{
\begin{array}{c}
\frac{3k_{d}}{2R}+\frac{2k_{c}(6-{c}_{0}R)}{R^{3}}<\frac{2k_{c}(10-{c}_{0}R)}{%
R^{3}},\quad (3k_{d}R^{2}<16k_{c}) \\
\frac{4\sqrt{3k_{d}k_{c}}}{R^{2}}+\frac{2k_{c}}{R^{3}}(2-{c}_{0}R),\quad
(3k_{d}R^{2}>16k_{c})
\end{array}%
\right.. \label{criticalpcmno}\eequ

Comparing equation (\ref{criticalpcm}) with (\ref{criticalpcmno}),
we find that in-plane modes have quantitive effect on the
stability of the cell membrane although they can not qualitatively
modify the result without considering them.

On the one hand, equation (\ref{criticalpcm}) includes the
classical results for stability of elastic shells. For classic
shell, the critical pressure for spherical shell is $\Delta
p_c\sim Yh^2/R^2$ \cite{landau,Pogorelov}, where $Y$ is the
Young's modulus of the shell. If taking $\bar{c}_0=0$, $k_d\sim
Yh$, $\bar{k}_c\sim Yh^3$ and $R\gg h$, our equation
(\ref{criticalpcm}) also gives $\Delta p_c\sim Yh^2/R^2$. On the
other hand, equation (\ref{criticalpcm}) includes the critical
pressure for stability of lipid membranes. For spherical lipid
membranes, the critical pressure is $\Delta p_c\sim k_c/R^3$
\cite{oy2}, which is the natural result of equation
(\ref{criticalpcm}) only if let $k_d=0$ in it. Otherwise, if we
take the typical parameters of cell membranes as $\bar{k}_c\sim
20k_B$T \cite{Duwe,Mutz2}, $k_d\sim 2.4\mu$N/m \cite{Lenormand},
$h\sim 4$nm, $R\sim 1\mu$m, $\bar{c}_0R\sim 1$, we obtain $\Delta
p_c\sim 2$ Pa from equation (\ref{criticalpcm}), which is much
larger than $\Delta p_c\sim 0.2$ Pa without considering $k_d$
induced by membrane cytoskeleton. This result reveals that
membrane cytoskeleton greatly enhances the mechanical stabilities
of cell membranes, at least for spherical shape.

\section*{CELL STRUCTURE DYNAMICS}
In above discussions, we only consider the static elasticities of
cell membranes. However, it is more important to understand the
dynamics of cells. At least, we much cover the dynamic behavior of
cell membrane, cytoplasm, cytoskeleton and nucleus. In this
section, we will propose a framework of cell structure dynamics
involving tensegrity architecture of cytoskeleton, fluid dynamics
of cytoplasm and elasticities of cell membranes with membrane
cytoskeleton. We make a model of a cell as follows.

 (i) Nucleus. The
nucleus is thought of as a small rigid sphere in the middle of
cell because it is a relative stiff membrane. Its inner structure
is neglected.

(ii) The tensegrity architecture of cytoskeleton
\cite{Ingber1,Ingber2}. Cytoskeleton is a cross-linked structure
comprised of microfilaments, microtubles and intermediate
filaments. Microfilaments are better at resisting tension while
microtubles at withstanding compression. Intermediate filaments
also resist tension, but only for significant strains of cells.
Thus the cytoskeleton is an integral system consists of continuum
tension elements (microfilaments) and discrete compression
elements (microtubles) prestressed by osmotic pressure acting on
the cell membranes and adhesions of other cell or extracellular
matrix. The total free energy of cytoskeleton might be written as
$F_{csk}=F_{ten}+F_{com}$, where $F_{ten}$ is free energy
contributed by tension elements and $F_{com}$ comes from
compression elements. $F_{ten}$ can be written as the production
of tension and total length of tension elements for large strains
because the tension is continuously transferred and the entropic
elasticity is not important. Its form for small strains must refer
to the entropic elasticity of biopolymers \cite{Dio}. $F_{com}$
depends on the stresses and strains in each compression elements.
In fact, $F_{csk}$ depends implicitly on the relative positions of
junction points between cytoskeleton and cell membrane. The
equilibrium conditions of force can be expressed as \bequ
\mathbf{f}_{i}=\partial F_{csk}/\partial
{\mathbf{r}_i}\label{intercsk}\eequ if we omit the inertial term
of cytoskeleton. Here \{$\mathbf{R}_i$\} and \{$\mathbf{f}_{i}$\}
represent the positions of junction points between cytoskeleton
and cell membrane, and the forces at that junction points induced
by cell membrane, respectively.

(iii) Cytoplasm and the liquid surroundings of the cell. Cytoplasm
and the liquid surroundings of the cell are regarded as
incompressible viscous fluid. The dynamics might be describe by
Navier-Stokes equation \cite{landaufl}: \beqn &&
\partial \mathbf{v}/\partial t+(\mathbf{v}\cdot
\mathbf{grad } )\mathbf{v}=-\mathbf{grad\ }p/\rho +(\eta/\rho)\mathrm{div\ }\mathbf{grad\ }\mathbf{v},\label{nseq1}\\
&&\mathrm{div\ } \mathbf{v}=0,\eeqn where $\mathbf{v}$, $\rho$,
$p$ and $\eta$ are the velocity vector, density, pressure and
dynamic viscosity of fluid, respectively. The components of
viscosity stress tensor $\bm{\tau}$ has the form  \bequ \tau_{ij}=
\eta(\partial v_i/\partial x_j+\partial v_j/\partial x_i),
(i,j=1,2,3),\eequ where $v_1$, $v_2$, and $v_3$ are, respectively,
three components of velocity in three Cartesian coordinates $x_1$,
$x_2$, and $x_3$.

(iv) The coupling between cell membrane and cytoskeleton,
cytoplasm as well as liquid surroundings. The equilibrium
equations of cell membrane under the interaction of cytoskeleton,
cytoplasm and liquid surroundings can be expressed as \beqn
&&\frac{k_d}{2}\left[(\epsilon _{22}-\epsilon _{11})\frac{\partial \sqrt{g_{22}}}{%
\partial u^{1}}-\sqrt{g_{22}}\frac{\partial }{\partial u^{1}}(2\epsilon
_{11}+\epsilon _{22})-\sqrt{g_{11}}\frac{\partial \epsilon _{12}}{%
\partial u^{2}}-2\epsilon _{12}\frac{\partial \sqrt{g_{11}}}{\partial
u^{2}}\right]\nonumber \\ &&\qquad=\mathbf{e}_1\cdot
\Delta\bm{\tau}\cdot \mathbf{e}_3+\sum_i
\mathbf{f}_i\cdot\mathbf{e}_1\delta(\mathbf{r}-\mathbf{r}_i), \label{membrdyn1}\\
&&\frac{k_d}{2}\left[(\epsilon _{11}-\epsilon _{22})\frac{\partial \sqrt{g_{11}}}{%
\partial u^{2}}-\sqrt{g_{11}}\frac{\partial }{\partial u^{2}}(\epsilon
_{11}+2\epsilon _{22})-\sqrt{g_{22}}\frac{\partial \epsilon _{12}}{%
\partial u^{1}}-2\epsilon _{12}\frac{\partial \sqrt{g_{22}}}{\partial
u^{1}}\right] \nonumber\\ &&\qquad=\mathbf{e}_2\cdot
\Delta\bm{\tau}\cdot \mathbf{e}_3+\sum_i
\mathbf{f}_i\cdot\mathbf{e}_2\delta(\mathbf{r}-\mathbf{r}_i), \\
&&\Delta p-2(\mu
+k_{d}J)H+\bar{k}_{c}(2H+\bar{c}_{0})(2H^{2}-\bar{c}_{0}H-2K)+\bar{k}_{c}\nabla
^{2}(2H) \nonumber\\
&&\qquad-\frac{k_{d}}{2}[a\epsilon _{11}+2b\epsilon
_{12}+c\epsilon _{22}] =\mathbf{e}_3\cdot \Delta\bm{\tau}\cdot
\mathbf{e}_3+\sum_i
\mathbf{f}_i\cdot\mathbf{e}_3\delta(\mathbf{r}-\mathbf{r}_i),\label{membrdyn}\eeqn
where $\Delta$ represent the outer quantity minus the inner one.
In above three equations, we omit the inertial term of cell
membrane because it is expected to be much smaller than the
viscosity force. Moreover, these equations are valid only for
small in-plane deformations of cell membranes. If we only
considering lipid membrane, these equations are degenerated to the
key equations in Ref. \cite{Komuras}.

Equations(\ref{intercsk})-(\ref{membrdyn}) are highly nonlinear
and coupling with each other so that they must be solved by
numerical methods. The key point is to develop a arithmetic to
dealing with the coupling boundary conditions
(\ref{membrdyn1})-(\ref{membrdyn}).

\section*{CONCLUSION}
In this chapter, we discuss elasticities and stabilities of lipid
membranes and cell membranes. We obtain the equations to describe
equilibrium shapes and strains of cell membranes by osmotic
pressures. We find that the critical pressure for spherical cell
membrane is much larger than that of spherical lipid bilayer
without considering membrane cytoskeleton. We also try to
construct a framework of cell structure dynamics involving
tensegrity architecture of cytoskeleton, fluid dynamics of
cytoplasm and elasticities of cell membranes with membrane
cytoskeleton. It is an important direction to develop arithmetic
to solve the coupling equations (\ref{intercsk})-(\ref{membrdyn})
in the future.


\begin{thebibliography}{99}
\bibitem{Lodish}Lodish, H. \textit{et al.} \textit{Molecular Cell
Biology}; W. H. Freeman \& Co.: New York, 1999; Chapter 1.
\bibitem{Edidin}Edidin, M. \textit{Nature Rev. Mol. Cell Bio.} 2003,
\textit{4}, 414-418.
\bibitem{nicolson}Singer, S. J.; Nicolson, G. L. \textit{Science} 1972, \textit{175},
720.
\bibitem{Helfrich}Helfrich, W. \textit{Z. Naturforsch. C} 1973, \textit{28},
693-703.
\bibitem{gennes}de Gennes, P. G. \textit{The Physics of Liquid Crystals}; Clarendon Press: Oxford,
1975; 100-103.
\bibitem{Duwe}Duwe, H. P.; Kaes, J.; Sackmann, E. \textit{J. Phys. Fr.} 1990, \textit{51},
945-962.
\bibitem{Mutz2}Mutz, M.; Helfrich, W. \textit{J. Phys. Fr.} 1990,
\textit{51}, 991-1002.
\bibitem{oy1}Ou-Yang, Z. C.; Liu, J. X.; Xie, Y. Z. \textit{Geometric Methods in the Elastic Theory of Membranes
in Liquid Cristal Phases}; World Scientific: Singapore, 1999.
\bibitem{Reinhard}Lipowsky, R. \textit{Nature} 1991, \textit{349},
475-481.
\bibitem{Seifertap}Seifert, U. \textit{Adv. Phys.} 1997, \textit{46},
13-137.
\bibitem{oy2}Ou-Yang, Z. C.; Helfrich, W. \PRL 1987, \textit{59},
2486-2489.
\bibitem{oy2pra}Ou-Yang, Z. C.; Helfrich, W. \PRA 1989, \textit{39}, 5280-5288.
\bibitem{oy4}Ou-Yang, Z. C. \PRA 1990, \textit{41}, 4517-4520.
\bibitem{Mutz}Mutz, M.; Bensimon, D. \PRA 1991, \textit{43},
4525-4527.
\bibitem{Linz}Lin, Z. \textit{et al.} \textit{Langmuir} 1994, \textit{10},
1008-1011.
\bibitem{Rudolph}Rudolph, A. S.; Ratna, B. R.; Kahn, B. \textit{Nature} 1991, \textit{352},
52-55.
\bibitem{fengyz}Feng, Y. Z. \textit{Bio-Mechanics}; Science Press: Beijing,
1983.
\bibitem{oy3}Naito, H.; Okuda, M.; Ou-Yang, Z. C. \PRE 1993, \textit{48},
2304-2307.
\bibitem{tzcjpa}Tu, Z. C.; Ou-Yang, Z. C. \JPA 2004, \textit{37}, 11407¨C11429.
\bibitem{tzcpre}Tu, Z. C.; Ou-Yang, Z. C. \PRE 2003, \textit{68}, 61915.
\bibitem{antu}An, R.; Tu, Z. C. \textit{Prerint} 2003, math-phys/0307007.
\bibitem{chen2}Chern, S. S.; Chen, W. H. \textit{Lectures on Differential
Geometry}; Peking University Press: Beijing, 2001; Chapter 6.
\bibitem{wangzx}Wang, Z. X.; Guo, D. R. \textit{Introduction to Special
Function}; Peking University Press: Beijing, 2000.
\bibitem{tuge}Tu, Z. C.; Ge, L. Q.; Li, J. B.; Ou-Yang, Z. C. \textit{Prerint} 2003,
cond-mat/0312319.
\bibitem{gennes2}de Gennes, P. G. \textit{Scaling Concepts in Polymer Physics}; Cornell University: New York, 1979.
\bibitem{Treloar}Treloar, L. R. G. \textit{The Physics of Rubber Elasticity}; Clarendon Press: Oxford, 1975.
\bibitem{wujk}Wu, J. K.; Wang, M. Z. \textit{Introduction to Elastic theory};Peking University Press: Beijing,
1981.
\bibitem{zhangz}Zhang, Z.; Davis, H. T.; Kroll, D. M. \PRE 1993, \textit{48}, R651-R654.
\bibitem{Pogorelov}Pogorelov, A. V. \textit{Bendings of surfaces and stability of
shells}; Providence, R.I.: AMS, 1989.
\bibitem{landau}Landau, L. D.; Lifshitz, E. M. \textit{Theory of Elasticity};
Butterworth-Heinemann: Oxford, 1997; 3rd edn.
\bibitem{Lenormand}Lenormand, G. \textit{et al.} \textit{Biophys. J.} 2001, \textit{81},
43-56.
\bibitem{Ingber1}Ingber, D. E. \JCS 2003, \textit{116}, 1157-1173.
\bibitem{Ingber2}Stamenovi\'{c}, D.; Ingber, D. E. \textit{Biomechan.
Model Mechanobiol.} 2002, \textit{1}, 95-98.
\bibitem{Dio}Dio, M.; Edwards, S. F. \textit{The Theory of Polymer
Dynamics}; Clarendon Press: Oxford, 1986.
\bibitem{landaufl}Landau, L. D.; Lifshitz, E. M. \textit{Fluid
Mechanics};Butterworth-Heinemann: Oxford, 1998; Chapter 2.
\bibitem{Komuras}Komura, S.; Seki, K. \textit{Physica A} 1993,
\textit{192}, 27-46.
\end{thebibliography}
\end{document}